\documentclass[aps,pre,showpacs,twocolumn,groupedaddress]{revtex4}
\usepackage{graphicx}
\def\be{\begin{equation}}
\def\ee{\end{equation}}
\begin{document}
\draft

\title{Superconducting pipes and levitating magnets}
\author{Yan Levin\footnote{levin@if.ufrgs.br} and Felipe B.
Rizzato\footnote{rizzato@if.ufrgs.br}}
\affiliation{Instituto de F\'{\i}sica,
Universidade Federal do Rio Grande do Sul\\
Caixa Postal 15051, 91501-970, Porto Alegre, RS, Brazil
}
\begin{abstract}
Motivated by a beautiful demonstration of the Faraday's 
and Lenz's law in which a small neodymium magnet falls 
slowly through a conducting
non-ferromagnetic tube, we  consider the dynamics of a magnet falling
through a superconducting pipe.  Unlike the case of normal conducting pipes,
in which the magnet quickly reaches the terminal velocity, inside a
superconducting tube the magnet falls freely. 
On the other hand, to enter the pipe the magnet must overcome a large 
electromagnetic energy
barrier.  For sufficiently strong magnets, the  barrier is so large that 
the magnet will not be able to penetrate it and will be
suspended over the front edge.  We calculate the work that must done 
to force the
magnet to enter a superconducting tube.  
The calculations show
that superconducting pipes are very efficient  at screening magnetic fields.
For example, the magnetic field of a dipole at the center of a short pipe of 
radius $a$ and length $L \approx a$ decays, in the axial direction, 
with a characteristic length 
$\xi \approx 0.26 a$.  The efficient screening of the magnetic field
might be useful for shielding highly sensitive superconducting quantum
interference devices, SQUIDs.  Finally, the motion of the magnet through a 
superconducting pipe is compared and contrasted to the flow of ions
through a trans-membrane channel.

\end{abstract}
%
\pacs{41.20.Gz, 74.25.Ha, 07.55.Nk}
\maketitle
%

\section{Introduction}

There is a beautiful demonstration of Faraday's and Lenz's
laws which became very popular as a result of an easy availability
of powerful rare earth magnets~\cite{sa92,we06}.  The demonstration consists
of a long pipe made of a conducting,
non-ferromagnetic material, such as copper or aluminum, and 
a neodymium magnet which is allowed to fall through it.  
One finds that the magnet
takes a very long time to traverse the pipe.  In fact for
a tube of about $2$m in length, the magnet takes almost
$25$s to finish the trip~\cite{we06}! On the other hand, a 
non-magnetic object of the
same dimensions falls through the  pipe in less than $1$s. 
It is quite amazing to observe the falling 
magnet from the top aperture,  the magnet appears to be moving through
a very dense fluid.  In reality,  air provides 
only a negligible resistance,
and what actually slows the magnet is the force 
produced by the
eddy currents induced in the pipe.  This force 
is proportional to the velocity of the falling
magnet.  When the drag force becomes equal to the magnet's weight,
acceleration  ceases and the fall
continues at a constant terminal velocity.  
For strong rare earth magnets, the terminal
velocity is reached very quickly.  Perhaps surprisingly, in view of the
complexity of the problem, it is actually possible to perform a fairly 
simple calculation which agrees quantitatively with the terminal velocity
observed experimentally~\cite{we06}. Curiously, the 
calculation also predicts 
that the terminal
velocity should be proportional to the electrical resistivity of the pipe's material.
This suggests that if the pipe is an ideal superconductor, 
the velocity of the falling magnet
should vanish.  One can easily see, however, that this  
conclusion can not be right. Suppose
that a magnetic dipole is created inside an infinite superconducting pipe.
During the process of creation, the magnetic field inside the pipe is 
changing and the electric current is induced on its surface.
The surface currents screen the magnet's
field and prevent it from entering into the interior of 
the superconductor.  
In the case of an ideal superconductor (inertialess electrons)
considered in the bulk of this paper, the
penetration length is zero and both the 
magnetic and the electric fields are perfectly screened.  
By symmetry it is also clear that for an infinitely long pipe,
the magnetic field produced by the induced currents 
is maximum precisely at the location of the magnet.  
Since the magnetic force 
on a dipole is proportional to the 
gradient of the field it must, therefore, vanish so that the  
magnet will fall without any resistance.   
The theory of
reference~\cite{we06} is not applicable to the perfect conductors
because  it was  explicitly constructed to treat normal metals 
for which the 
magnetic permeability is very close to that of vacuum.  Furthermore, the rate
of decay of the induced currents in such metals 
is very fast, compared to the magnet fall
velocity, allowing us to neglect the effects of self-induction~\cite{we06}.  
Clearly, neither one of these conditions is met in the case of
super or ideal conductors which
dynamically screen  magnetic field from their interior. 
As the resistivity of the pipe metal 
is decreased, there will be a crossover from the terminal  
velocity found in ref. \cite{we06} for normal pipes 
to the free fall velocity inside perfectly conducting pipes.

Although a magnet ``created'' in the interior of an infinite 
superconducting pipe
will fall freely under the action of the gravitational field, it
takes work to  bring (create) the magnet inside the pipe 
in the first place. This is so because the magnetic field lines, which for a 
free dipole spread throughout the space, must now be confined  
in a much more restricted volume.  In this paper we will calculate
the work that must be done to bring a magnet into 
superconducting pipe of length $L$ and radius $a$. 
Furthermore, the formalism developed here can be easily extended 
to study more general problems of screening of the magnetic
field in cylindrical geometry~\cite{BaBe64,ThGi76}, which are of particular
interest for the development of reliable SQUID devices~\cite{clay99}.

\section{The model}

\begin{figure} []
\includegraphics[scale=0.5]{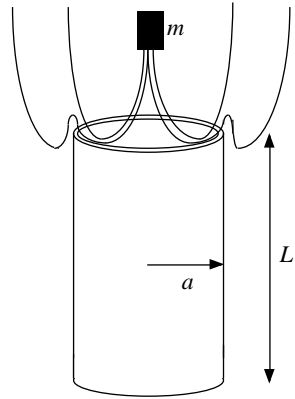}
\caption{Schematics representation of a magnetic dipole descending into a 
superconducting tube. The flux lines are repelled from the tube.\label{schematics.eps}}
\end{figure}

The model that we shall study 
is depicted in  Fig. \ref{schematics.eps}. A 
magnetic dipole, of moment ${\bf m} = m \hat {\bf z}$,  
is brought from infinity and is inserted into cylindrical 
superconducting  pipe
of length $L$ and radius $a$ along the symmetry axis. 
This axis is taken to coincide with
the $z$ axis of the coordinate system. The radius of the pipe 
is assumed to be
much larger than the wall thickness  and will, therefore,
be ignored.  The Faraday's law of induction 
requires that
\begin{equation}
\frac{d \Phi(z,t)}{dt}=-\oint {\bf
E}\cdot d{\bf r},
\label{0}
\end{equation}
where  $\Phi(z,t)$ is the magnetic flux passing through a cross section of the
pipe at position $z$ and time $t$, and ${\bf E}$ is the local electric field.
Since the tangential component of the electric field
is continuous across the superconductor/air interface,
the right hand side of
Eq.~(\ref{0}) must vanish because no electric field can 
be present inside a 
perfect conductor.  Thus, the flux passing
through any cross section of the pipe must be constant in time.
Furthermore, when
the magnet is at infinity, the flux entering the pipe is 
zero, which then means that 
$\Phi(z,t)=0$ at all future times as well. 
Vanishing flux is a direct
consequence of the physics of cylindrical superconductors and must be used as a
boundary condition for the solution of the Maxwell's equations. 
In Fig. 1 we have made an attempt to represent the magnetic field lines 
which are not allowed to go through 
the pipe because of the restriction on flux. Again we stress that
this behavior is very different from what happens with  
normal metals~\cite{we06} for which the
electric field does not vanish and magnetic flux changes
through different 
cross sections of the pipe.
 
Inside the superconductor,
magnetic field is zero and the continuity of the normal component of 
${\bf B}$ requires that the normal component of the external field 
also vanishes at the superconducting wall.  The
magnetic field lines must, therefore, be tangent to the pipe's surface.  
In general,
however, vanishing of ${\bf B}_n$ at the interface is 
not sufficient to fully specify the boundary 
condition necessary for the existence of a unique solution
to Maxwell equations. 

\section{Uniqueness}

What are the boundary conditions which will make the field equation
\begin{equation}
\nabla \times {\bf B} = \mu_0 {\bf J} 
\label{equa1}
\end{equation}
have a unique solution?  
Suppose that Eq.~(\ref{equa1}) allows for two distinct solutions
${\bf B}_1$ and ${\bf B}_2$ for the {\it same} value of the current
density ${\bf J}$.  
Since $\nabla \cdot {\bf B}=0$, we can always define 
a vector potential ${\bf A}$
such that ${\bf B}_{1,2} = \nabla \times {\bf A}_{1,2}$. 
Now, consider the integral over all space
\begin{equation}
\int \delta {\bf B} \cdot \delta {\bf B} \> d^3r = \int \nabla \times \delta {\bf A}
\cdot \nabla \times \delta {\bf A} \> d^3 r,
\label{equa2}
\end{equation}
where $\delta {\bf B} \equiv
{\bf B}_2-{\bf B}_1$ and $\delta {\bf A} \equiv
{\bf A}_2-{\bf A}_1$
The integrand on the right-hand-side can be cast into more 
convenient form using
the identity
$(\nabla \times \delta {\bf A})^2 \equiv \delta {\bf A} \cdot \nabla \times \nabla \times
\delta {\bf A} +\nabla \cdot (\delta {\bf A} \times \nabla \times \delta {\bf
A})$.   Recalling that the current sources for the fields {\it 1} and {\it 2} are 
identical: $0 = \mu_0 ({\bf J}_2 - {\bf J}_1) \equiv \mu_0 \, \delta {\bf J} = \nabla
\times
\delta {\bf B} =
\nabla
\times \nabla
\times \delta {\bf A}$, the integrand of Eq. (\ref{equa2}) reduces to a perfect 
divergence.  Using the divergence theorem, the volume integral can now be transformed
into an integral over the bounding surfaces $S$, which in our case are the 
cylinder and the spherical shell of radius $R=\infty$.  
\begin{equation}
\int \delta B^2 \> d^3 r = \int_S da \> \hat {\bf n} \cdot (\delta{\bf A} \times
\nabla \times \delta{\bf A}).
\label{equa3}
\end{equation}
Since the dipolar field decreases rapidly with distance, the contribution
to the integral in Eq.~(\ref{equa3}) coming from the spherical shell at $R=\infty$ 
vanishes, and $S$ reduces to the surface
of the cylinder.
Using the cyclical invariance of the integrand, we see that 
Eq.~(\ref{equa1}) has a unique solution 
($\delta {\bf B}({\bf r}) \equiv  0$) either if 
$\hat{\bf n} \times \delta{\bf A}=0$ or 
$\hat{\bf n} \times \delta{\bf B}=0$, {\it on the pipe surface}.  
To have a well posed problem it is, therefore, 
not sufficient to specify only the normal component of the 
magnetic field, 
instead the tangential components
of the field at the interface must be provided.  In a cylindrical geometry the
boundary conditions posed in terms of the vector potential are particularly
useful. Azimuthal symmetry around the 
$z$ axis, restricts the vector potential to have only one non zero component
in the ${\bf e}_{\phi}$ direction, 
${\bf A} = A_{\phi} \, \hat {\bf e}_{\phi}(r,z)$. 
The flux through a cross section of the pipe  is then
\begin{equation}
\Phi(z) = \int {\bf
B} \cdot \hat{\bf n} da = \int \nabla \times {\bf A} \cdot \hat{\bf n} da = 
\oint {\bf A} \cdot \, d{\bf
r} = 2 \pi a\,A_{\phi}(a,z).
\label{equa0}
\end{equation}
Specification of the flux passing through the pipe is, therefore, equivalent
to the specification of $A_{\phi}$  and guarantees that 
Eq.~(\ref{equa1}) has a unique solution in the cylindrical geometry. 
  
\section{A pipe of infinite length}

For superconducting pipes of $L=\infty$, Eq.~(\ref{equa1}) 
can be solved analytically, while 
for finite $L$ only numerical solution is possible.  
We start, therefore, with the  $L=\infty$ case.

The dipole of moment ${\bf m} =
m \hat {\bf z}$ is located inside the pipe on the axis of symmetry 
at $z=0$ $r=0$. 
In the pipe's interior there are no
free currents and the Ampere's law, Eq.~(\ref{equa1}), reduces to 
$\nabla \times {\bf B}=0$ outside the magnet.  The magnetic field can then be 
written as a gradient of a scalar function ${\bf B} = - \nabla
\varphi$.  This  defines the scalar magnetic potential $\varphi$, which 
also satisfies
the  Laplace equation since $\nabla \cdot {\bf B}=0$.
The magnetic potential  $\varphi \equiv  \varphi_d +
\varphi_{ind}$ is produced by the point dipole, 
\begin{equation}
\varphi_d={\mu_0\,m z \over 4 \pi (r^2+z^2)^{3/2}},
\label{equa4}
\end{equation}  
and by the currents induced on the surface of the superconductor.  Since 
Eq.~(\ref{equa4}) is a solution of the Laplace equation, so must be 
$\varphi_{ind}$,
\begin{equation}
{1 \over r}\, {\partial \over \partial r} r {\partial \over \partial r} \varphi_{ind} + 
\partial_z^2 \varphi_{ind} = 0,
\label{equa6}
\end{equation}
for $r<a$.
We next note that  $\varphi_{ind}$ must be odd in $z$, free of singularities, 
and must vanish as $z \rightarrow \infty$.  Under these conditions, the
solution of the Laplace equation (\ref{equa6}) can be written in terms of
a Fourier integral involving modified Bessel function of the
first kind,
\begin{equation}
\varphi_{ind}(r,z) = {2 \over \pi} \, \int_0^\infty A(k) I_o(k r) \sin(k z) dk.
\label{equa8}
\end{equation}
According to the results of the previous section, the function $A(k)$ will be uniquely 
determined by the condition that $\Phi(z)=0$,  which means that 
no magnetic field lines are 
lost to the wall.  For infinite superconducting pipe this boundary condition,
is, therefore, equivalent to the vanishing of the normal component of the   
magnetic field $B_n(a,z) = 0$ on the wall. This means that the 
normal component of the induced field
must  cancel  {\it exactly} the field produced by the dipole, 
\begin{eqnarray}
\partial_r \varphi_{ind}|_{r=a} = -\partial_r \varphi_d|_{r=a} 
= \frac{3 \mu_0 m z a} {4 \pi (a^2+z^2)^{5/2}},
\label{equa5}
\end{eqnarray} 
at the air/superconductor interface.
Using the standard identities, Eq.~(\ref{equa5}) can now be written in 
terms of the
modified Bessel function of the second kind $K_1(x)$,
\begin{eqnarray}
\partial_r \varphi_{ind}(r,z)\,|_{r=a} = \cr 
{\mu_0 m \over 2\pi^2} \, \int_0^\infty k^2 K_1(k a) \sin (k z) dk, 
\label{equa9}
\end{eqnarray}
%
Combining expressions (\ref{equa8})
and (\ref{equa9}) and using 
$dI_0(x)/dx = I_1(x)$,  
enables us to calculate the function
\begin{eqnarray}
A(k) = \frac{\mu_0 \, m}{4 \pi} \frac{k K_1(k a)}{I_1(k a)} \;.
\label{equa9a}
\end{eqnarray}
The magnetostatic scalar potential inside an infinite pipe is then
\begin{widetext}

\begin{equation}
\varphi(r,z) = {\mu_0\,m z \over 4 \pi (r^2+z^2)^{3/2}}+\frac{ \mu_0 \,m}{ 2 \pi^2} \int_0^{+\infty} dk\,k \sin(k z) {K_1(k a) \over I_1(k a)}  I_0(kr)\;.
\label{equa10}
\end{equation}

\end{widetext}
The first
term on the right hand side of Eq.~(\ref{equa10}) is the potential
produced by the point dipole located at $r=0, z=0$, while the second
term is the magnetostatic scalar potential produced by the electric
currents induced on the superconducting surface.  
For large $z$, Eq.~(\ref{equa10}) 
simplifies to
%
%
%
\begin{equation}
\varphi(r,z) \approx \frac{\mu_0 m |z|}{2 \pi a^2 z}\left(1  + 
{J_0(k_1 r/a)\over J_0(k_1)^2}\, e^{- k_1 |z|}\right)\,,
\label{equa13}
\end{equation}
where $k_1$ is the first root of the Bessel function 
$J_1$: $J_1(k_1)=0$ with $k_1 \approx 3.831$. 
The axial magnetic field of a dipole inside
a superconducting pipe, 
\begin{equation}
B_z(r,z) \approx \frac{\mu_0 m }{2 \pi a^2} 
{k_1 J_0(k_1 r/a)\over J_0(k_1)^2}\, e^{- k_1 |z|}\,,
\label{equa13c}
\end{equation}
is, therefore, strongly screened, with a characteristic
length  $\xi = a/k_1 \approx 0.26 a$. 

To confine a magnet inside a superconducting tube of small radius costs a lot
of energy.  The  work necessary to achieve this
can be calculated using a charging process,
\begin{equation}
W=-\int_0^1 d \lambda \, {\bf m}\cdot {\bf B}_{ind}(\lambda{\bf m})=-
{1\over 2} {\bf m}\cdot {\bf B}_{ind}({\bf m}), 
\label{equa13a}
\end{equation}
in which the dipole is charged from $0$ to its final value $m$ while
the induced field ${\bf B}_{ind}$ 
responds accordingly.  Using Eq.~(\ref{equa10}) we find
\begin{equation}
W=\frac{\mu_0 \, m^2}{2 \pi^2 a^3} \int_0^\infty dx \,x^2 \frac{K_1(x)}{I_1(x)} \approx  
0.797 \frac{\mu_0 \, m^2}{4 \pi a^3} \;.
\label{equa13b}
\end{equation}

The condition that the 
magnetic field lines  must
exit the pipe on the same side on which they entered ($\Phi(z)=0$), 
implies that the magnet can not
``probe'' the full length of the pipe.  Thus, if the pipe length is such that $L>a$,
its field screening properties should be identical to those of a pipe of $L=\infty$.  
With this observation in mind we are now ready to study superconducting
pipes of finite length.

\section{Pipes of finite lengths} 
\label{finite}

Consider a superconducting pipe of finite length $L$ placed 
along the $z$ axis, whose center coincides with the origin at  
$z=0$. Results of the previous sections 
suggests that if $L > a$, 
shortly after the dipole finds itself inside the tube, 
the magnetic field configuration 
should be identical to that inside an infinite pipe  
and the axial force should vanish. However, little can be said about what kind of 
forces act upon the dipole as it enters or exits the pipe. Finite 
length case should, therefore, be examined with some care. 

From sections \S 2 and \S 3 we recall that the magnetic flux at any cross
section of a superconducting  pipe must vanish.  
This information will be central to
the practical aspects of the theory  from now on. As the dipole approaches the
pipe from $z \rightarrow +\infty$, surface  currents are generated over the
pipe's  surface. It is convenient to imagine
that the length of the pipe is subdivided into uniform rings, each  carrying
a circulating surface current density $j(z)$.  The vector potential produced by
these currents has only $\phi$ component and its magnitude is 
given by the linear superposition,
\begin{equation}
A_{ind} (r,z) = \frac{a \mu_0}{2} \int_{-L/2}^{+L/2} K(r,z,z') j(z') dz',
\label{equa13p5}
\end{equation}
where the kernel
\begin{equation} 
K(r,z,z')=\int_0^\infty dk e^{-k |z-z'|} J_1(k r) J_1(k a) 
\label{equa14a}
\end{equation}
is obtained from 
the field produced by one thin ring \cite{jack75}. 
The flux generated by the surface currents   
at $z$ is $\Phi_{ind}(z) = 2 \pi a A_{ind} (a,z)$,
see Eq. (\ref{equa0}).  For $r=a$ the kernel can be evaluates explicitly 
in terms of the hypergeometric function, 
\begin{equation}
K(a,z,z') = {a^2  \,_2 F_1 [3/2,3/2;3;-{4 a^2 \over |z-z'|^2}] \over 2  |z-z'|^3}.
\label{equa14}
\end{equation}
Condition that through each cross section of the pipe the net 
flux vanishes, $\Phi(z,t)=0$,
leads to an integral equation for the surface currents,
\begin{equation}
\int_{-L/2}^{+L/2} K(z,z') j(z') dz'=
-\frac{m }{2 \pi \left[(z_m-z)^2+a^2\right]^{3/2}},
\label{equa14b}
\end{equation}
where $z_m$ is the coordinate of dipole.
To determine $j(z)$, Eq.~(\ref{equa14b}) is solved numerically by first 
discretizing the integral and then performing a matrix inversion. 
%
%
Once the current distribution is known, the vector 
potential $A_{ind}$ is calculated using Eq.~(\ref{equa13p5}). Knowing $A_{ind}$, 
the induced
magnetic field on the 
axis of symmetry $r=0$,  ${\bf B}_{ind} = {\bf \hat z}/r\,
\partial_r (r A_{ind})|_{r=0}$ and  
the magnetic force on the dipole
${\bf F} = {\bf \hat z} \partial_z ({\bf m}\cdot{\bf B}_{ind})|_{z=z_m,r=0}$ 
can be easily evaluated. We note that forces arising from 
flux trapping in type-II superconductors are neglected in the present discussion. 
For this to be a good approximation,
the superconductor must have either 
high critical state current  density $J_c$ or 
be subjected to only small magnetic 
field~\cite{davis90,schilling04,campbell71}. We will discuss  this 
more in  conclusions.

\subsection{Checking the accuracy of the numerical procedure} 

\begin{figure} [ht]
\includegraphics[scale=0.5]{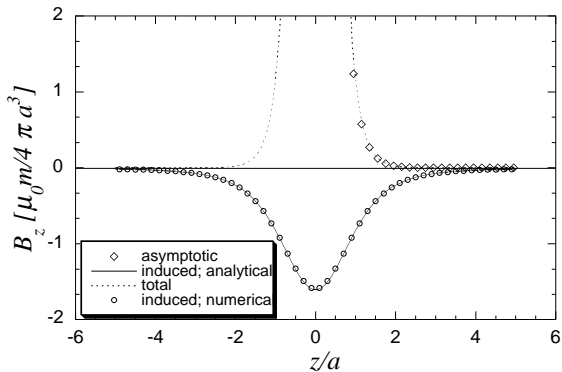}
\caption{Comparison of the numerical (circles) and the 
analytical (full line) 
calculations for the induced magnetic field.  The analytical result
is for $L=\infty$, while the numerical solution is for $L=10 a$.
Nevertheless, there is a perfect agreement between the two.  
We also compare
the analytical asymptotic form (dotted line) Eq.~(\ref{equa13c}) 
of the axial magnetic field, with the result of 
numerical integration (squares) for $L=10 a$,  showing a clear  
exponential decay of the axial field, even inside a 
finite superconducting pipe.
\label{analXnum.eps}}
\end{figure}

The accuracy of the numerical procedure used to solve the 
integral equation Eq.~(\ref{equa14b}) 
can be judged by comparing it  with the
analytical solution for an infinite pipe, Eq.~(\ref{equa10}). 
In Fig. \ref{analXnum.eps} we compare $B_{ind}$ along 
the axis of symmetry for an infinitely long pipe with the field
calculated using a numerical integration of  Eq.~(\ref{equa14b}) 
for a pipe of  $L=10 a$ with a dipole located at $z_m=0$. 
The agreement is perfect and attests to the reliability of the numerical
solution. Furthermore, since the axial magnetic field is maximum at the
position of the dipole, the force on it will vanish, in agreement with our
previous discussion.

In Fig. \ref{analXnum.eps} we also show that asymptotically the magnetic
field is well approximated by Eq.~(\ref{equa13c}). Thus, in agreement with the
previous discussion, the axial magnetic field 
is screened exponentially even for pipes of
finite length, as long as $L>a$.  

\subsection{Edge effects for superconducting tubes of finite lengths}

\begin{figure}
\includegraphics[scale=0.5]{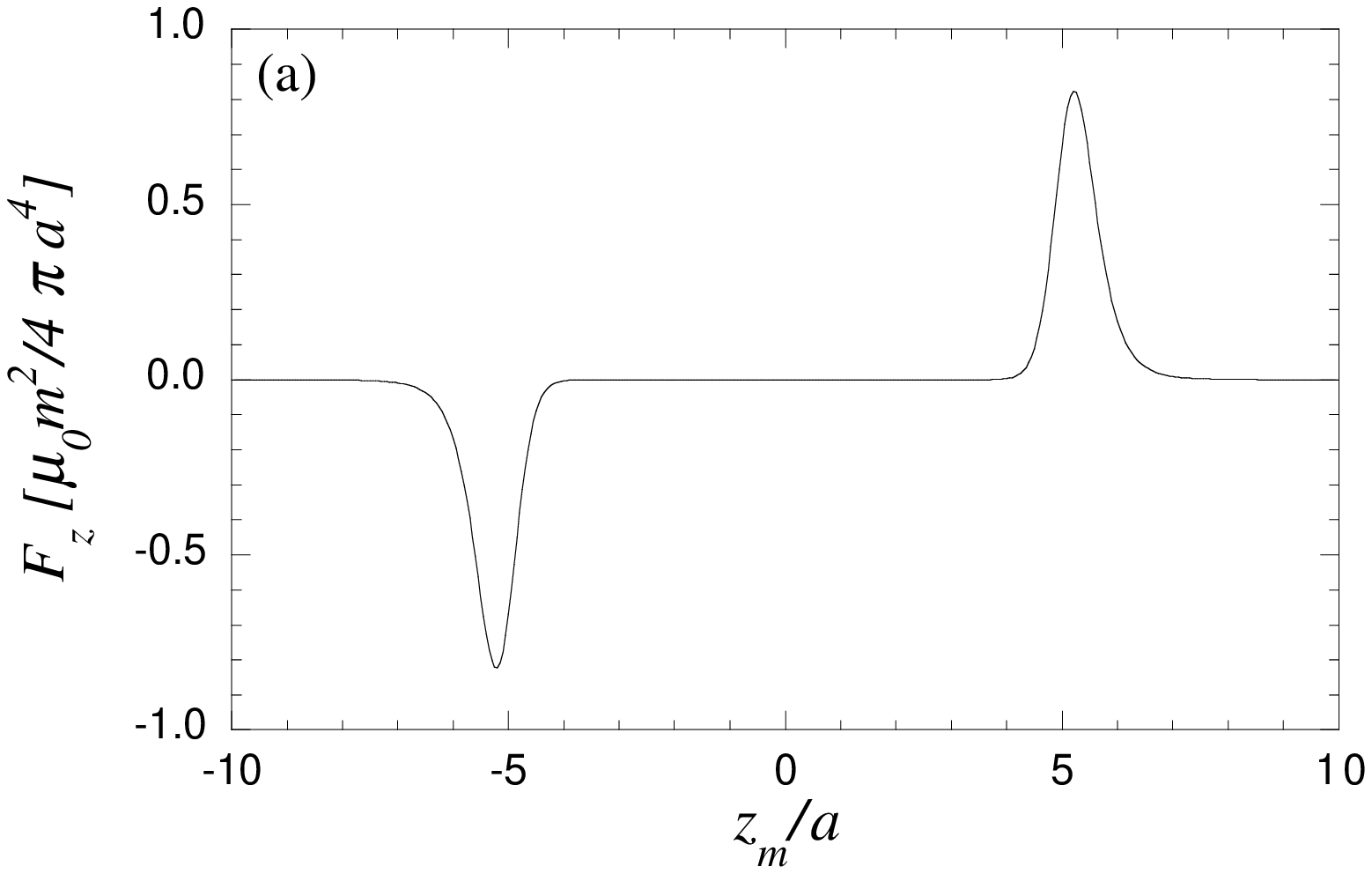}

\includegraphics[scale=0.5]{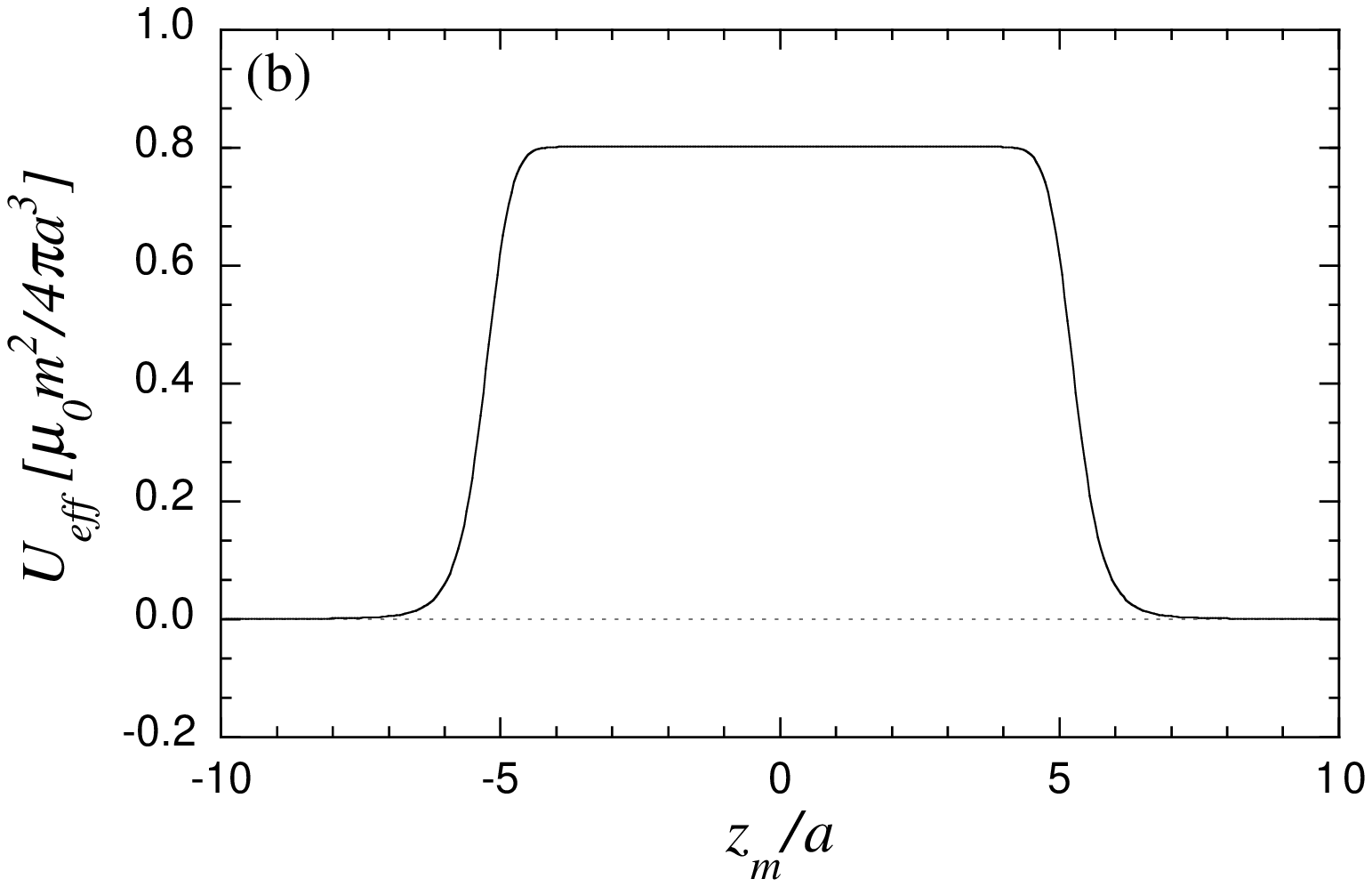}
\caption{Force acting on a magnetic dipole and the associated 
potential energy as it moves into and out of a superconducting pipe. 
Negative force is repulsive. 
Although, the force peaks look symmetric, in reality
the force decays algebraically outside the tube entrance, and exponentially into
the tube. 
\label{forca.eps}}
\end{figure}

Satisfied with the accuracy of the numerical procedure, we can now use it
to study the edge effects associated with the finite length  
superconducting pipes. In Fig. \ref{forca.eps}
we plot the magnetic force
felt by a magnet as it moves from infinity into the interior of a 
superconducting pipe of length $L=10 a$. Panel (a)
shows  a strong repulsive force near the pipe entrance 
which vanishes rapidly as the magnet penetrates into the pipe.
In the case of small neodymium magnets (weight $6$g) used in our 
demonstrations of the Faraday's
and Lenz's laws~\cite{we06} and a superconducting pipe of radius 
$a=7.85\,mm$ we find the repulsive force to be sufficient to support
a weight of $1\,kg$! 
In panel (b) we plot the work necessary to bring a magnet from infinity to
a point $z$.  Clearly, there is a large
electromagnetic energy barrier that the magnet must overcome to enter the pipe. 
Note that soon after the magnet crosses the front edge into the interior of
the tube, the potential stabilizes at a plateau the value of which 
is in perfect agreement with the work found to be necessary to confine a dipole
inside an {\it infinite} superconducting tube, Eq.~(\ref{equa13b}). This
agreement is, again, a consequence of a very efficient exponential 
screening of the magnetic field by the surface currents, even inside  pipes of
{\it finite length}.

\begin{figure} [h]
\includegraphics[scale=0.5]{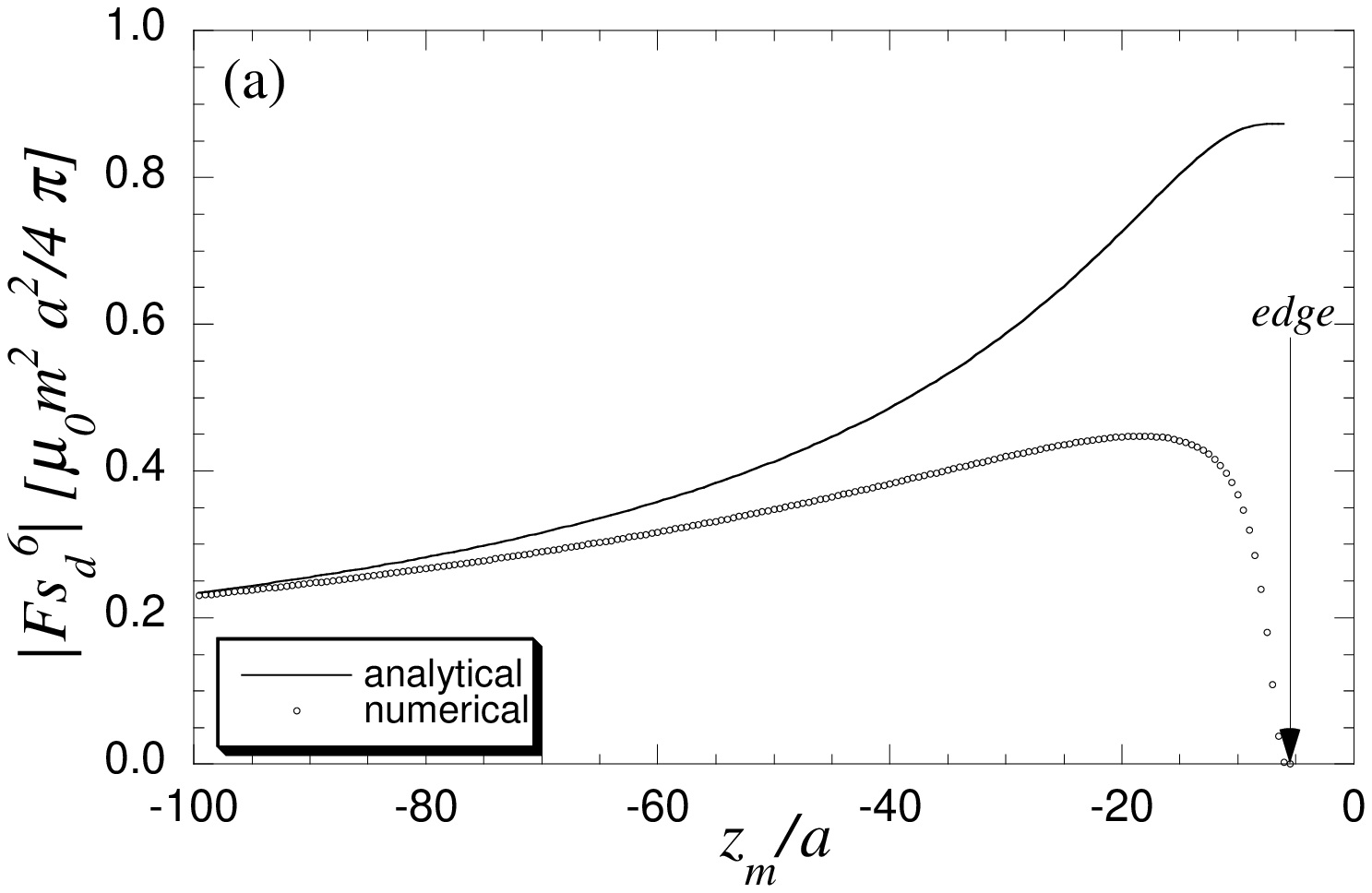}

\includegraphics[scale=0.5]{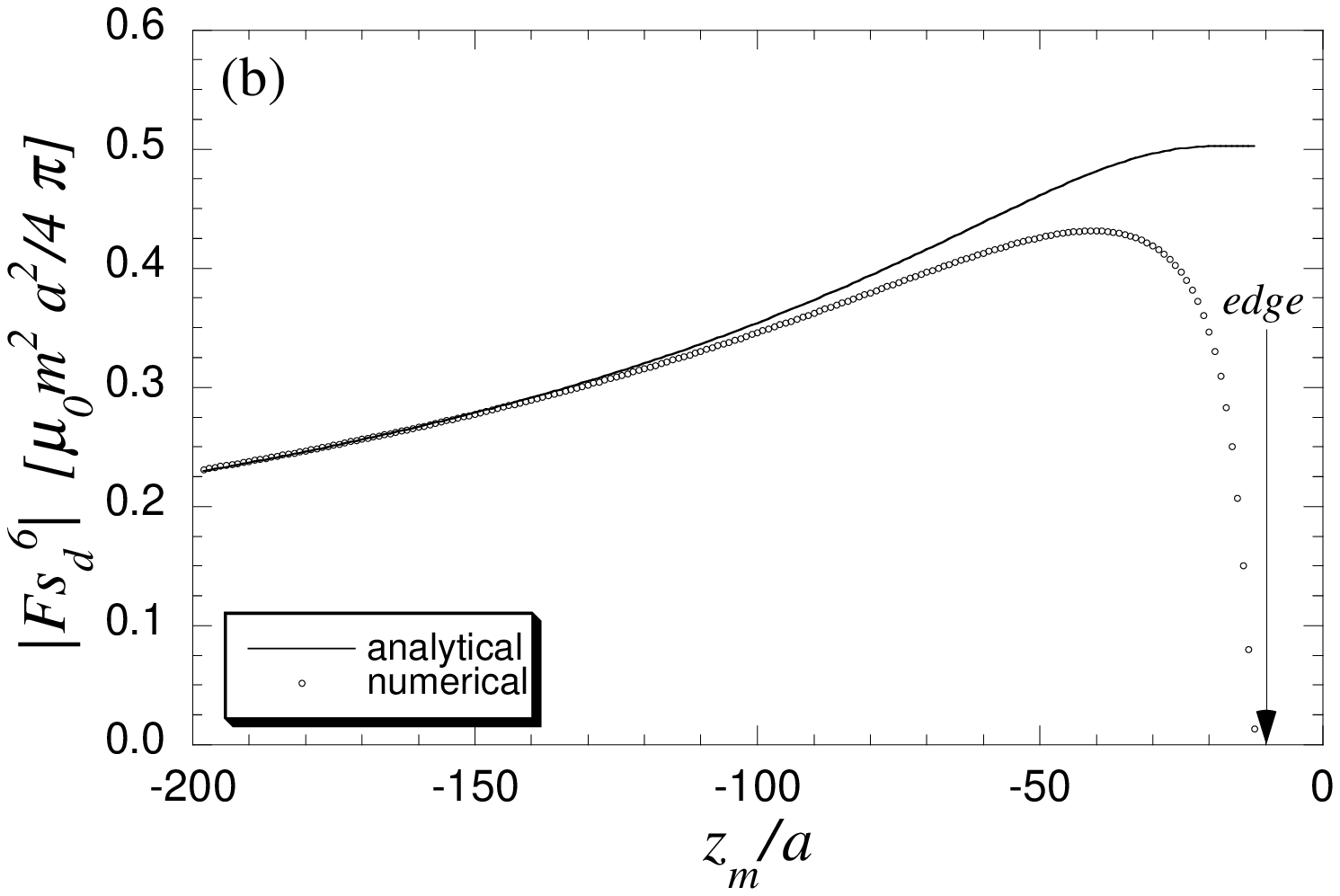}
\caption{Asymptotic force felt by a dipole when it is 
far from a superconducting pipe.  Points are the result
of the numerical calculation and the solid lines are the estimates
obtained using Eq.~(\ref{equa16}). Panel (a) is for $L=10 a$
and panel (b) is for $L=20 a$. \label{eles.eps}}
\end{figure}
As the magnet approaches a superconducting tube, 
it experiences a magnetic force which opposes its motion. The
magnitude of this force can be estimated using a simple scaling argument.
Magnetic field produced by a dipole decays with distance as 
$B_{magnet} \sim m/s^3$. If the superconducting pipe is 
sufficiently narrow ($L \gg a$) 
and the magnet is not too close to the front
edge, the mutual-inductance  effects
between the sections of the pipe can be neglected and the
induced surface current density at distance $s$ from the magnet will be
$j(s) \sim B_{magnet} \sim m/s^3$.  On the other hand, each one of these
current loops will produce a magnetic field at the position of the
dipole, 
$d B_{pipe \rightarrow magnet}  \sim j(s) \,ds / s^3 \sim m\, d s/s^6$,
and will result in a net force
\begin{equation}
|F_z| = \alpha m^2\int_{s_d}^{s_d+L} ds/s^7 = {\alpha m^2 \over s_d^6} f\left({s_d \over L}\right),
\label{equa16}
\end{equation}
where $s_d=z_m-L/2$ is the distance of the magnet to the front edge of the pipe
and $\alpha$ is the pipe's
polarizability factor. The scaling function is found to be  
$f(x) \equiv 1-x^6/(1+x)^6$. 
Asymptotically, the force $F_z$ behaves as 
\begin{equation}
|F_z s_d^6| \rightarrow \cases{ \alpha m^2\,\,\, {\rm if} \,\,\, a \ll |s_d| \ll L, \cr
\left({6 \alpha m^2 L \over z_m}\right) \,\,\, {\rm if} \,\,\, |s_d| \gg L.}
\label{equa17}
\end{equation}
Thus, if $L \gg a$, the force on a magnet 
decays algebraically outside the pipe 
and vanishes exponentially after the magnet
enters the pipe.
 
The polarizability $\alpha$ can be calculated  by considering
the far asymptotic limit of Eq.~(\ref{equa17}).  Under this condition, the
magnetic field varies only slightly over the length of the pipe, and the
induced current can be calculated analytically by 
solving Eq.~(\ref{equa14b}).  We find,
\begin{equation}
I=\frac{m L}{2 \pi z_m^3}.
\label{equa17a}
\end{equation}
This  current will produce a magnetic field at the position of  the dipole 
resulting in a repulsive force 
\begin{equation}
F=\frac{3 \mu_0 m^2 a^2 L }{4 \pi z_m^7}.
\label{equa17b}
\end{equation}
Comparing Eqs.~(\ref{equa17b}) and (\ref{equa17}) we find the polarizability 
of the superconducting pipe to be $\alpha=\mu_0 a^2/ 8 \pi$.

%
%

In Fig. \ref{eles.eps}(a) we  compare the asymptotic force calculated using the numerical
solution of the integral equation ~(\ref{equa14b}) with the estimate
obtained using Eq.
(\ref{equa16}) for a tube of $L = 10 a$. 
The agreement is very good for
$z_m \gg L$, but poorer for smaller values of the coordinate. 
The problem is that
$L = 10 a$ is not sufficiently  large to well satisfy the
inequality $a \ll z_m \ll L$, as demanded by the first asymptotic region of
(\ref{equa17}). For larger values of $L$, such as
$L=20$ of Fig. \ref{eles.eps}(b), the agreement  already is much better.  
The abrupt decay of $F s_d^6$ near the 
entrance of the tube is a consequence  of 
the  finite cross section of the pipe, which becomes important for small values
of $s_d$. 

\section{Conclusions}

We have examined the forces and the fields produced by a  
small magnet as it enters into a  
superconducting pipe.
For an infinitely long pipe, magnetic field produced by 
a magnet inside a tube is exponentially screened by the 
surface currents.  

In the case of finite superconducting pipes 
we find that the magnetic field is also exponentially
screened, as long as the length of the pipe is larger than its diameter.  
The exponential screening of the magnetic field
is a consequence of the Faraday's law which leads to 
vanishing of the  magnetic flux inside a 
superconducting pipe.   The flux lines  
can not go through the pipe and 
at any cross section the number of lines going down is the 
same as the number of lines coming up.   
The electromagnetic cost of confining a dipole
inside a superconducting tube is, therefore,  very large,  
since the field lines must be strongly
compressed to fit inside the pipe. 
We calculate that in order to insert a  small neodymium
magnet of $6$g into a superconducting pipe of radius $8$mm requires
$10$N of force. Once the magnet is inside the pipe, however, its
motion will continue unhindered.  
At the exit, the magnet will be
ejected with the same force that was required to insert it into the 
pipe in the first place. 

It is curious to compare the magnet's motion through a superconducting pipe,
with the flow of ions through an ion channel~\cite{Le06}. 
Ion channels are water filled holes, 
responsible for a small potential gradient that exists across all biological 
membranes. Since the dielectric constant inside a pore is much
larger than the dielectric constant of a phospholipid membrane,  
the normal component of the {\it electric field} at the
channel/membrane interface is very small and  
the electric field lines are mostly confined to stay 
in the pore's interior.  In this respect, the pore is very similar
to a superconducting pipe, for which the normal component of the {\it magnetic field}
also vanishes on the boundary.  
Nevertheless, one finds that contrary to what happens to a magnet inside 
a superconductor,
the repulsive force on an ion  
does not vanish even when it is far inside the channel. 
Ions always feel a force 
which tries to expel them from the channel 
(except exactly at midpoint where, by symmetry, the force vanishes).  

The difference between superconducting pipes and ion channels
is precisely due to the additional constraint of 
vanishing flux imposed
by the Faraday's law on a perfect conductor.  The requirement that
the flux through any cross section of a superconducting pipe 
must vanish, forces the system to be in a metastable
state.  If this condition is relaxed, energy of the magnet/superconductor 
system can  be lowered.  For example,  suppose that 
the magnet is placed inside a {\it normal} conductor
and only {\it later} the temperature is lowered until the superconducting state is achieved.
In this case the magnetic field configuration inside the pipe will be 
different from the one when 
the magnet is placed into an already superconducting pipe. 
There will no longer be a restriction that the flux must vanish and, in fact,
the magnetic field lines at the moment that the pipe turns  superconducting
will be partially expelled, leaving behind a finite flux.  
The energy cost of confining  a magnet  under these conditions will, 
therefore, be significantly lower. Inside an ion channel, 
behavior of the electric field  is 
very similar to the case
of a superconductor without a vanishing flux constraint.  It is precisely 
this constraint that is responsible for the exponential screening of the 
magnetic field found for superconducting pipes of even finite
length.  If the vanishing flux constraint is relaxed (by say the process of turning
the pipe superconducting {\it after} the magnet is placed inside), the 
magnetic field will no longer be exponentially screened inside a {\it finite} pipe,
and the magnet would encounter a repulsive barrier, similar to the one faced by an ion
inside a trans-membrane channel, and a non-vanishing magnetic force~\cite{Le06} 
      
In this work we have neglected the
hysterectic ( flux trapping) forces typically present in type-II superconductors. 
For small cylindrical neodymium magnets weighing $6$g and radius  
$r =6.35 \, mm$ moving through a superconducting pipe of 
$a \sim 8 \, mm$,  the magnetic field at the pipe surface is
on the order of $B \sim 0.1 \, T$.  Considering a  strong 
pinning condition with a high critical state 
current density $J_c \sim 10^{8} A/m^2$, we 
estimate~\cite{davis90,schilling04,bean64}  the drag force on the magnet 
to be only a  small fraction $ \sim 0.0001$ of the
force required to enter the superconducting pipe. 
Thus,  our results  should not be much
affected even if more realistic type II 
superconductors are used.

\acknowledgments

This work is partially supported by CNPq, Brazil. We would like to thank 
Renato Pakter, Paulo Pureur, Osvaldo Schilling and 
Fernando L. da Silveira for interesting discussions.

\end{document}